\documentclass{article}[12pt]
\usepackage{amsmath,amssymb,latexsym,amsthm,amscd,amsxtra}
\usepackage[all]{xy}
\usepackage{epsfig}
\usepackage{graphicx}
\usepackage[bottom]{footmisc}
\usepackage[section]{placeins}

\def\baselinestretch{2.0}
\def\nn{\nonumber}

\begin{document}

\begin{center}
{\bf \Large Unintended consequences of imprecise Notation  \\-  an example from mechanics}

\vspace*{0.25in}
{\bf \Large Asim Gangopadhyaya\footnote{agangop@luc.edu} and Gordon Ramsey\footnote{gramsey@luc.edu}}

\vspace*{0.25in}
{\large \it Department of Physics, Loyola University Chicago, \\1032 W. Sheridan Rd., Chicago, IL 60660, U.S.A.}
\end{center}

\vspace*{0.25in}
\noindent
{\begin{abstract}
We present a conundrum that results from the imprecise use of notation for partial derivatives. Taking an example from mechanics, we show that lack of proper care in representing partial derivatives in Lagrangian and Hamiltonian formulations paradoxically leads to two different values for the time derivative of the canonical momentum. This problem also exists in other areas of physics, such as thermodynamics. \end{abstract}}

\def\baselinestretch{2.0}
\vspace*{0.25in}

\vspace*{0.25in}

\vspace*{0.25in}
We generally insist that our students speak and write with precision, but they often do not. One case in point is that of partial differentiation, an operation that has a ubiquitous presence in many areas of physics, especially in areas of thermodynamics and canonical transformations. When we represent the partial differentiation of a function $\phi\left(x_1, x_2, \cdots, x_n\right)$ of several independent variables with respect to one of those variables $x_k$ by $\frac{\partial\phi}{\partial x_k}$, we often do not underscore the constraints in this variation; i.e., the variables that are kept constant while ${x_k}$ changes. As the following example from classical mechanics shows, this omission can frequently lead to serious misunderstanding.

The comparative study of the Lagrangian and Hamiltonian approaches to classical mechanics is a standard topic in undergraduate courses in mechanics. The crux of the difference between the two approaches is that they have two different sets of independent variables.  This aspect  is not always sufficiently appreciated by students. Here we present a simple paradox that challenges the reader to engage deeply, and  thus develop a clearer perspective on each formalism, and their differences. Paradoxes provide many benefits in instruction, namely to motivate deeper thinking and provide a more thorough understanding of the topic. As pointed out by Welch \cite{paradox} in this journal, their greatest benefit is to address ``specific deficiencies in understanding and reasoning."  We shall see that the resolution of our paradox lies in the careful handling of partial differentiation.  Hence, one of our aims is to urge students to be cognizant of the roles played by various variables in the variation of a function that depends on more than one physical quantity.

In undergraduate classical mechanics courses, students are generally encouraged to express the Lagrangian, $\cal L$ and Hamiltonian, $\cal H$ in various coordinate systems. This not only helps them build very important expertise in choosing an efficient set of generalized coordinates for a system, often using symmetries, it also helps them explore how $\cal L$ and  $\cal H$ could look very different in one coordinate system compared to another, and yet carry exactly the same physical information.

Let us consider a Lagrangian of a free system; i.e., the potential $V(q)=0$. The Lagrangian is given by
\begin{eqnarray}
 {\cal L}(q,\dot{q}) = T(q,\dot{q}) - V(q)  = T(q,\dot{q})
\end{eqnarray}
The canonical momentum is defined by
\begin{eqnarray}
p \equiv  \frac{\partial  {\cal L}}{\partial \dot{q}} ~.
\end{eqnarray}
Then, the Lagrange's equation of motion $\frac{d}{dt}\, \frac{\partial  {\cal L}}{\partial \dot{q}} = \frac{\partial {\cal L}}{\partial {q}}$ implies
\begin{eqnarray}
\dot p   = \frac{\partial T}{\partial {q}} \label{Lag-implication}~.
\end{eqnarray}
Let us do a similar analysis using Hamiltonian formulation. Assuming the kinetic energy to be a homogeneous quadratic function of velocities, and since the potential $V$ is identically zero, the Hamiltonian is  $T+V = T$ \cite{Marion,Goldstein}. Hamilton's equation of motion then implies,
\begin{equation}
\dot p   = -\frac{\partial  {\cal H}}{\partial {q}}  = -\frac{\partial T}{\partial {q}}.
\end{equation}
Thus, the conundrum is that $\dot p$ is given by ${\frac{\partial T}{\partial {q}} }$ in the  Lagrangian formulation, and by  $-{\frac{\partial T}{\partial {q}} }$ in the Hamiltonian formalism; i.e.,
\begin{eqnarray}
\dot p   = \left.{\frac{\partial T}{\partial {q}} }\right|_{\rm Lagrangian\, Formulation}
= - \left.{\frac{\partial T}{\partial {q}} }\right|_{\rm Hamiltonian\, Formulation}
\label{conundrum}~.
\end{eqnarray}
This suggests that  $\dot p =0$, as one would expect in a case of a free particle.  Is this necessarily true? For example, a particle described by cylindrical coordinates ($r, \phi, z$), the kinetic energy is given by  $T=\frac12 m \dot{r}^2+\frac12 m r^2\dot{\phi}^2 +\frac12 m \dot{z}^2$, the derivative $\frac{\partial T}{\partial {r}}\neq 0$, and hence $\dot p_r \neq 0$. Thus, for this case, it becomes difficult to explain how a non-zero $\dot p_r$ be equal to $\frac{\partial T}{\partial {r}}$ in Lagrangian formulation and $-\, \frac{\partial T}{\partial {r}}$ in Hamiltonan formulation.

Similarly, for a particle described by spherical coordinates ($r, \theta,\phi$), the kinetic energy is given by
$T=\frac12 m \dot{r}^2+\frac12 m r^2\dot{\theta}^2+\frac12 m r^2\sin^2\theta~\dot{\phi}^2$, neither the
$\dot p_r $, nor the $\dot p_\theta$ are equal to zero. So, how do we explain that $\dot p_\theta$ is equal
to $\frac{\partial T}{\partial {\theta}}$ in the Lagrangian formulation and $-\frac{\partial T}{\partial {\theta}}$
in the Hamiltonian approach, and yet $\dot p_\theta\neq 0$?  As we will show later, the solution for this apparent
problem lies in the fact that  $\frac{\partial T}{\partial {q_i}}$ calculated in two different formulations
are different.  It is crucial to know which variables are being kept constant in these partial differentiations.
For the Lagrangian formulation, the derivative $\frac{\partial T}{\partial {q_i}}$ implies that $T$ is being differentiated, keeping coordinates $q_j$ with $j\neq i$ constant, and all $\dot q_j$'s are held constant. For the Hamiltonian formulation, the derivative $\frac{\partial T}{\partial {q_i}}$ implies that $T$ is being differentiated keeping coordinates $q_j$ with $j\neq i$ constant, and all $p_j$'s  are held constant.  Since the constraints under which these two derivatives are being taken are different, the derivatives themselves may be exactly equal in magnitude and opposite in sign, and hence may make
Eq. (\ref{conundrum}) consistent with a non-zero $\dot p$.

Let us check this possibility with a free particle in spherical coordinates. The Lagrangian and the Hamiltonian of the system are given by ${\cal L}=\frac12 m \dot{r}^2+\frac12 m r^2\dot{\theta}^2+\frac12 m r^2\sin^2\!\theta\,\dot{\phi}^2$ and
${\cal H}=\frac{p_r^2}{2m}+\frac{p_\theta^2}{2mr^2}+\frac{p_\phi^2}{2mr^2\sin^2\theta}$ respectively \cite{books}. From Lagrange's equation, we get
\begin{eqnarray}
\dot{p_\theta} =\left.\frac{\partial T}{\partial {\theta}}\right|_{\{r,\phi,\dot r, \dot \theta, \dot \phi\}} = m r^2\sin\theta\,\cos\theta\,\dot{\phi}^2~.\label{p_theta_Lag}
\end{eqnarray}
From Hamilton's method, we get
\begin{eqnarray}
\dot{\phi}=\frac{p_\phi}{mr^2\sin^2\theta}\label{p_theta_Ham}~; ~{\rm and}~~
\dot{p_\theta} =-\left.\frac{\partial T}{\partial {\theta}}\right|_{\{r,\phi,p_r, p_\theta, p_\phi\}} =
\frac{p_\phi^2\cos\theta}{mr^2\sin^3\theta}.
\end{eqnarray}
Substituting for $p_\phi$ in Eq. (\ref{p_theta_Ham}), we see that $\dot{p_\theta}$ is indeed equal to $m r^2\sin\theta\,\cos\theta\,\dot{\phi}^2$ in both cases.

Thus, in spherical coordinates, Eq. (\ref{conundrum}) does not lead to any inconsistency for $\dot{p_\theta}$. We now show that this is indeed true in all coordinate systems, provided that the kinetic energy is a homogeneous quadratic function of the generalized velocities.

Let us start with a system with zero potential \footnote{It is worth noting that this discrepancy in the values for $\dot p$ returned by the two formalisms is present even for the case with a non-zero potential $V(r,\theta,\phi)$; it just appears more dramatic when the potential is zero.} and with $N$-degrees of freedom, described by the Lagrangian
\begin{eqnarray}
 {\cal L}(q_i,\dot{q}_i,t)= \frac12\sum_{ij}^N F_{ij}(q)\dot{q}_{i}\dot{q}_j~, \label{Lagrangian}
\end{eqnarray}
where the \underline{symmetric} coefficients $F_{ij}$ are, in general, functions of $N$-independent coordinates $q_i$. From here on, we will suppress the $q$-dependence of $F_{ij}$. The canonical momenta $p_i$ are given by
\begin{eqnarray}
p_i = \frac{\partial  {\cal L}(q_i,\dot{q}_i,t)}{\partial \dot{q}_i} = \sum_{j} F_{ij}\,\dot{q}_j \equiv F_{ij}\,\dot{q}_j , \label{momenta}
\end{eqnarray}
where summation over repeated indices is assumed. Since coordinates $q_i$ are assumed to be independent, the inverse of the matrix $ F_{ij}$ exists, and allows us to invert Eq. (\ref{momenta})  to find velocities in terms of momenta. We denote the elements of this inverse matrix by  $\left(F^{-1}\right)_{ij}$.
From  Eq. (\ref{momenta}), solving for the velocities, we get
\begin{eqnarray}
\dot{q}_i  =  \left(F^{-1}\right)_{ij}\, p_j~.\label{velocities}
\end{eqnarray}
Since the Hamiltonian must be written as a function of $q_i, p_i$ and $t$, the Hamiltonian for the system is then given by
\begin{eqnarray}
{\cal H}(q_i, p_i,t) &=& p_i\dot{q}_{i}-{\cal L} = p_i\dot{q}_{i} - \frac12 F_{ij}\dot{q}_{i}\dot{q}_j~\label{ten}\\
&=& p_i\left(F^{-1}\right)_{ij}\, p_j - \frac12 F_{ij}
\left(  \left(F^{-1}\right)_{ik}\, p_k\right)
\left(  \left(F^{-1}\right)_{j\ell}\,  p_\ell\right) ~,
\end{eqnarray}
where we have substituted velocities from Eq. (\ref{velocities}) into Eq. (\ref{ten}).
Now, using the associativity of matrix multiplication, we get
\begin{eqnarray}
{\cal H}(q_i, p_i,t)
&=& p_i\left(F^{-1}\right)_{ij}\, p_j - \frac12 \underbrace{\left( F_{ij}
 \left(F^{-1}\right)_{ik}\right)}_{\delta_{jk}}\, p_k\,
\left(  \left(F^{-1}\right)_{j\ell}\,  p_\ell\right) \nn~\\
&=&  p_i\left(F^{-1}\right)_{ij}\, p_j - \frac12 \, p_k\,
\left(  \left(F^{-1}\right)_{k\ell}\,  p_\ell\right) \nn\\
&=&  \frac12 ~p_i\left(F^{-1}\right)_{ij}\, p_j ~,\label{Hamiltonian}
\end{eqnarray}
where we have relabeled the dummy indices in the second term of the second line.
Since the matrix $F_{ij}$ is a function of $q_i$'s, the Hamiltonian has been properly represented in terms of $q_i$, $p_i$ and t.
Now, let us compute $\dot p_i$ for this general case using the two different formalisms.
In the Lagrangian formalism,  from Eq. (\ref{Lagrangian}), we have
\begin{eqnarray}
\dot p_i   = {\frac{\partial \cal L}{\partial {q_i}} } =  {\frac{\partial T}{\partial {q_i}} } =
\frac12 \left[\frac{\partial}{\partial q_i} F_{jk}\right]\dot{q}_{j}\dot{q}_k
\label{Lag2}~.
\end{eqnarray}
On the other hand, from Eq. (\ref{Hamiltonian}),  the Hamiltonian formalism gives
\begin{eqnarray}
\dot p_i   = -{\frac{\partial \cal H}{\partial {q_i}} } =  -{\frac{\partial T}{\partial {q_i}} } = -
\frac12 \left( \frac{\partial }{\partial q_i}  \left(F^{-1}\right)_{\ell m}  \right)~p_\ell\,\, p_m
\end{eqnarray}
Now, substituting for momenta from Eq. (\ref{momenta}), we get
\begin{eqnarray}
\dot p_i   &=& -{\frac{\partial T}{\partial {q_i}} } = -
\frac12 \left( \frac{\partial }{\partial q_i}  \left(F^{-1}\right)_{\ell m}  \right)~
\left( F_{\ell j} \,\dot{q}_j\right)
\,
\left( F_{mk} \,\dot{q}_k\right)\nn~\\
&=&
\frac12 ~\left[ -F_{j\ell} \left( \frac{\partial }{\partial q_i}  \left(F^{-1}\right)_{\ell m}  \right)~F_{mk}\right]
\,\dot{q}_j
\,
\,\dot{q}_k
\label{Ham2}~.
\end{eqnarray}
In Eqs. (\ref{Lag2}) and (\ref{Ham2}), we have computed the time derivatives of momentum $p_i$ from Lagrangian and Hamiltonian formalisms respectively.  Consistency of these two expressions requires that factors within square brackets of these two equations be equal; i.e.,
\begin{eqnarray}
\left[\frac{\partial }{\partial q_i} F_{jk} \right]=\left[ -F_{j\ell} \left( \frac{\partial }{\partial q_i}  \left(F^{-1}\right)_{\ell m}  \right)\,F_{mk}\right]
\label{consistency1}~.
\end{eqnarray}
Multiplying the above equation by $ \left(F^{-1}\right)_{sj}$ and summing over the index $j$, we get
\begin{eqnarray}
\left(F^{-1}\right)_{sj}\,\left[\frac{\partial }{\partial q_i} F_{jk} \right] &=& \left(F^{-1}\right)_{sj}\,\left[ -F_{j\ell} \left( \frac{\partial }{\partial q_i}  \left(F^{-1}\right)_{\ell m}  \right)\,F_{mk}\right] \nn\\
&=& - \underbrace{  \left(F^{-1}\right)_{sj}\,F_{j\ell}}_{\delta_{s\ell}} \left( \frac{\partial }{\partial q_i}  \left(F^{-1}\right)_{\ell m}  \right)\,F_{mk}\nn\\
&=& -  \left( \frac{\partial }{\partial q_i}  \left(F^{-1}\right)_{sm}  \right)\,F_{mk}
\nn\\
&=& -  \left( \frac{\partial }{\partial q_i}  \left(F^{-1}\right)_{sj}  \right)\,F_{jk}\label{consistency2}~.
\end{eqnarray}
Thus, consistency of the two results requires
\begin{eqnarray}
\left(F^{-1}\right)_{sj}\,\left[\frac{\partial }{\partial q_i} F_{jk} \right]+ \left( \frac{\partial }{\partial q_i}  \left(F^{-1}\right)_{sj}  \right)\,F_{jk}=0
\label{consistency3}~.
\end{eqnarray}
However, the left-hand side of Eq. (\ref{consistency3}) is a perfect differential that can be written as
\begin{eqnarray}
\frac{\partial }{\partial q_i}      \left(  \left(    F^{-1}\right)_{sj}  \, F_{jk} \right)= \frac{\partial }{\partial q_i}      \delta_{sk}
\label{consistency4}~,
\end{eqnarray}
which is identically zero. Thus, for a system where  the kinetic energy is a homogeneous quadratic function of the velocities, the two seemingly different expressions for $\dot{p}_i$, one derived from the Lagrangian and the other from the Hamiltonian formulation, are thus proved to be identical.

\noindent
{\bf Conclusion:} In this note, we present a conundrum that appears in coordinate systems where the kinetic energy
is a quadratic homogeneous function of velocities and a function of generalized coordinates. The conundrum is most clear in the absence of an interaction, as both the Lagrangian and the Hamiltonian are equal to the kinetic energy, and the expressions for $\dot p_i$ in two formalisms are $\frac{\partial T}{\partial q_i}$ and $-\frac{\partial T}{\partial q_i}$ respectively. The purpose of this note is to show that if $\cal L$ and $\cal H$ are expressed correctly in their respective sets of independent variables, and the
differentiations are done carefully, then there is no discrepancy. This type of puzzle goes a long way toward instilling a deeper understanding of the partial differentiations involved in these two different formulations of
classical mechanics. We would like to stress that this problem is not localized to the area of mechanics alone. There are many examples in other fields as well. For example, in thermodynamics, the molar specific heat of a gas $\frac{\Delta Q}{\Delta T}$ could refer to either $C_p$ or $C_v$ depending on whether pressure or the volume is kept constant during the variation of temperature. Similarly, as pointed out by one of the referees, for an ideal gas we have  $dU=\frac32 n\, R\, dT= \frac32 p\, dV$. Thus, while $\left. \frac{\partial U}{\partial V} \right|_p = \frac32 p$, $\left. \frac{\partial U}{\partial V} \right|_T = 0$.  These examples drawn from different areas of physics clearly show that for partial differentiation, we need to pay great deal of attention to the variables that are constrained to a constant value.

We would like to thank the anonymous referees and Professor Thomas Ruubel for a very careful reading of the manuscript and for providing helpful suggestions. One of us (AG) would like to thank Alpana for her inspiration, and we both would like to thank Aysel Bayrak and Sureyya Terrace, Troya, Istanbul for the warm hospitality where this work was completed.


\end{document}